# Self-Assembly of Nanoparticles from Evaporating Sessile Droplets: Fresh Look into the Role of Particle/Substrate Interaction

N. Bridonneau,[1,2] M. Zhao,[1] N. Battaglini,[2] G. Mattana,[2] V. Thévenet,[1] V. Noël,[2] M. Roché,[1] S. Zrig[2*] and F. Carn[1*]

[1] Université de Paris, Laboratoire Matière et Systèmes Complexes, CNRS, UMR 7057, Paris, France.
[2] Université de Paris, ITODYS, CNRS, UMR 7086, 15 rue J-A de Baïf, F-75013 Paris, France

[*] E-mail : florent.carn@univ-paris-diderot.fr, samia.zrig@univ-paris-diderot.fr

**Abstract**

We studied the dependence of solid deposit shape obtained by free drying of sessile drops on the particles concentration and Derjaguin–Landau–Verwey–Overbeek (DLVO) particle/substrate interaction. In contrast to previous contributions using pH as a control parameter of interactions, we investigated an unprecedentedly wide range of concentrations and particle/substrate DLVO forces by modifying the nature of the substrate and particles as well as their size and surface chemistry whereas long-distance repulsive interactions between particles were maintained for most of the drying time. Our main result is that the different shapes of deposits obtained by modifying the particle concentration are the same in the different regimes of concentration regardless of particle/substrate interaction in the studied range of DLVO forces and particle concentrations. The second result is that, contrary to expectations, the dominant morphology of dry patterns at low particle concentration always shows a dot-like pattern for all the studied systems.

# Introduction

The drying of a drop of particle dispersion in open air on a solid substrate generates the formation of a solid deposit (i.e. dry pattern) which can take different morphologies.[1] These morphologies are influenced by several parameters such as particles concentration, particles and substrate wettability, temperature (i.e. atmospheric, substrate), relative humidity, particle "interactions" (i.e. particle/substrate, particle/particle, particle/liquid-gas interface) to name a few.[1–4] Understanding the role of each parameter on pattern formation is important because this evaporative process is at the basis of a large number of coating applications[5] with a renewed activity linked, in part, to the development of inkjet printing methods that enable the fast elaboration of nanostructures over a wide variety of substrates.[6–8] The purpose of the present work is to reconsider the role of the particle/substrate interaction on the shape of final deposits when long distance repulsive Derjaguin–Landau–Verwey–Overbeek (DLVO) particle/particle interactions are maintained for most of the drying time. This aspect which has been little studied[1,9–15] so far might become important with the emergence of (i) plasmonic applications requiring deposits with nanoscale resolution[16,17] and no structural defects, such as cracks; and (ii) biomedical applications linking diagnosis to the shape of a deposit obtained by drying a sample of biofluid.[18–21]

The dynamic of particles self-assembly at the contact line during evaporation was studied at the local scale by Yan et al.[22]. They showed that attractive DLVO interaction between polystyrene microparticles (~ 800 nm in size) and a negatively charged glass substrate leads to a low mobility of the particles in the vicinity of the substrate resulting in disordered local packing, regardless of the overall shape of the final deposit. This observation shades previous results obtained by Denkov et al.[23] showing that the dynamic of two-dimensional ordering is governed by the capillary force and the convective flow, while the DLVO force between particles play no significant role. The role of the DLVO force on the macroscopic shape of the solid deposit was also studied experimentally and numerically. Bhardwaj et al.[11] considered dispersions of $TiO_2$ nanoparticles (~ 25 nm in size) deposited on a glass substrate. They varied the particle/substrate and particle/particle DLVO forces by modifying the pH of the dispersions. They concluded that the deposit shape is dictated by the competition among (i) the evaporation-driven flow favoring ring-shaped deposits; (ii) the DLVO interaction between particles and substrate favoring uniform deposits when attractive; and (iii) the Marangoni flow, favoring a central bump deposit with a diameter much smaller than the initial drop diameter. The respective influence of particle/particle and particle/substrate interactions on the morphology of the dry pattern was also studied theoretically by Zigelman and Manor in 2017.[24] Their model accounts for an irreversible particle adsorption onto the substrate by using the boundary-layer theory and for irreversible coagulation of particles in the bulk using the Smoluchowski relation. They notably predict the formation of uniform deposits when the contribution of adsorption dominates that of coagulation. They underline the agreement with Bhardwaj et al.'s observations performed at low pH where coagulation is negligible due to repulsive particle/particle interactions. Overall, Anyfantakis and Baigl underline in their review,[10] that, regardless of their origin (i.e. DLVO or hydrophobic), most of existing experimental results agree on the fact that attractive particle/substrate forces promote the formation of uniform deposits. However, in the case of DLVO forces, we point out that this statement is based on few experimental studies[11,12] conducted on a single particle type at a fixed concentration using pH as a mean of controlling interactions. A disadvantage of this approach is that pH controls, at the same time, particle/substrate and particle/particle interactions, or even, the affinity of the particles for the liquid-gas interface. This multiple role complicates the analysis of one single contribution among others in the pH range around the particle isoelectric point.

Looking more generally at experiments conducted under particular drying conditions, or free-drying experiments where the role of the DLVO forces has also been described but indirectly without explicitly stating so, the effect of the DLVO forces seems unclear. For instance, Moraila-Martínez et al.[13] studied the role of particle/particle and particle/substrate interactions on the deposit formation without macroscopic evaporation using the "controlled shrinking sessile drop" method which enable to control the contact line dynamic. They considered different particles, substrates and pH for a fixed particle concentration. They observed ring-shaped deposits in all cases and they could not correlate the characteristic lengths of the deposits to the sign and magnitude of substrate/particle interactions. They concluded that the final morphology of the deposits would be driven by the particle/particle interaction rather than the particle/substrate interaction. Lee et al. studied the effects of particle size, concentrations of nanofluids and nature of the substrates on the residue patterns formed after drying $Al_2O_3$ and $TiO_2$ aqueous nanofluid droplets.[25] They obtained ring-shaped patterns at low particle concentrations and uniform patterns at high concentrations whatever the systems except for substrates with low wettability.

In this context, we used an alternative strategy to further explore the potential role of particle/substrate DLVO forces for freely evaporating drops. It consisted in considering model substrates and particles of different nature, surface chemistry and size (in the case of particles), whereas long-distance repulsive interactions between particles were maintained at fixed pH.

This method enabled us to cover a range of DLVO forces 3 orders of magnitude higher than that investigated in previous studies. For each set of particles and substrate, we considered at least 8 particle concentrations, over 4 decades, to identify the effect of interactions in the different concentration domains.

Dry deposits were observed with a scanning electron microscope equipped with a field emission gun (i.e. SEM-FEG) to probe dry pattern shapes and structures on length scales ranging from macroscale to nanoscale. Our main result is that the different shapes of deposits obtained by modifying the particle concentration are the same in the different regimes of concentration regardless of particle/substrate interaction in the studied range of DLVO forces and particle concentrations. The second result is that, contrary to expectations, the dominant morphology of dry patterns at low particle concentration exhibits dot-like pattern for all the studied systems.

## Experimental Section

**Materials.** Gold (III) chloride trihydrate ($HAuCl_4 \cdot 3H_2O$, > 99.99 %), trisodium citrate dihydrate ($Na_3C_6H_5O_7 \cdot 2H_2O$, ≥ 99 %), cysteamine ($HS(CH_2)_2NH_2$, noted CEA, 95 %), 11-mercaptoundecanoic acid ($HS(CH_2)_{10}CO_2H$, noted MUA, 98 %), α-lipoic acid ($C_8H_{14}O_2S_2$, ≥ 99 %) and silicon wafers ⟨111⟩ were purchased from Sigma-Aldrich and used as received without further purification. Aluminia coated silica nanoparticles Klebosol® 30CAL50 (noted $SiNP^+$ in the following) were purchased from Merck. Silica nanoparticles Snowtex® ST-OL (noted $SiNP^-$ in the following) were a gift from Nissan Chemical Industries Ltd., Tokyo, Japan. All the content of a gold salt powder batch was used at the first opening to prepare, using a glass spatula, a mother solution at 10 g/L in milliQ water that was stocked for periods not exceeding 3 months in a dark area to minimize photo-induced oxidation. The same batch of trisodium citrate has been used for all syntheses; it has been stored in desiccators after first opening. All glassware and teflon-coated magnetic bars were washed thoroughly with freshly prepared aqua regia and rinsed with milliQ water after each synthesis. All solutions were prepared with milliQ water (R = 18.2 MΩ).

**Synthesis of Gold Nanoparticles.** The citrate stabilized gold nanoparticles were synthesized by Turkevich reaction, following a reported procedure.[26,27] Briefly, 97 mg of trisodium citrate (0.33 mmol) were put in 150 mL of water (2.2 mM solution) and refluxed for 15 min, before adding 1mL of a 10 g/L solution of $HAuCl_4 \cdot 3H_2O$. The solution was kept on reflux for 10 min, then the heating was stopped and the solution was slowly cooled down to 90 °C, leaving the reaction mixture in the oil bath. The seeds had their size increased by repeating several growing cycles as followed: 55 mL of the solution were withdrawn, followed by the addition of 53 mL of water and 1mL of a 60 mM citrate solution. As soon as the temperature reached 90 °C again, 1 mL of $HAuCl_4.3H_2O$ were added and the solution was stirred for 30 min. Then, another portion of 1 mL of $HAuCl_4.3H_2O$ was added and the solution was stirred again for 30 min. These growing cycles were repeated up to the desired size of AuNPs. Typically, 3 cycles were required to obtain 25 nm (diameter) particles.

Citrate replacement by 11-mercaptoundecanoic acid (MUA) was achieved in two steps by adapting a protocol from literature.[28] Typically, 15 mL of AuNP(citrates) were put in a vial and 0.30 mL of a 0.5 M KOH solution were added. Then, 1.5 mL of α-lipoic acid (10 mM solution in ethanol) were added and the reaction was stirred overnight. The mixture was centrifuged (15000 rpm, 30 min) and the clear supernatant was discarded. The gold nanoparticles were resuspended in 15 mL of milliQ water and 0.30 mL of 0.5 M KOH were added, followed by 1.5 mL of MUA (30 mg in ethanol). The solution was stirred overnight, then centrifuged again. The supernatant was discarded and the AuNP(MUA) were resuspended in milliQ water to reach the desired concentration and kept in solution, in the dark. The UV-vis spectrum showed a slight widening of the SPR band upon ligand exchange associated with a 3 nm shift. The nanoparticles were concentrated by ultracentrifugation up to a final concentration of $2.46 \times 10^{11}$ $NP.mL^{-1}$ ($4.08 \times 10^{-9}$ $mol.L^{-1}$).

**Gold Substrate Preparation.** Gold surfaces were prepared in the clean room by the successive deposition on silicon wafers of a 5 nm-thick layer of titanium followed by a 200 nm-thick layer of gold, using a Corial D250 PECVD (Plasma Enhanced Chemical Vapor Deposition) system. The surfaces were annealed by butane flame to ensure a good crystallinity of the topmost layers, then rinsed successively in a bath of milliQ water (10 min) then absolute ethanol (10 min) and nitrogen dried. The functionalization of surfaces was carried out using a standard protocol.[29–31] The positively charged substrates were obtained by immersion of the surfaces in an ethanolic solution of cysteamine (10 mM) for 3h, whereas the negatively charged substrates were immerged in an aqueous solution of MUA (10 mM) for 12h. The positive (respectively negative) substrates were then sonicated in ethanol (resp. water) to desorb the non-grafted molecules and rinsed in ethanol then water before being dried under nitrogen flow. RMS roughness was typically of 1.6 nm for the gold surfaces, and 2.8 nm for MUA or CEA-gold surfaces (see AFM images 15x15µm² in Figure S3).

**Drop Deposition and Drying.** Deposition of the nanoparticles was done by careful dropping 0.5 µL of each solution on a horizontal substrate using a manual micropipette or an automated syringe pump (i.e. Krüss DSA 100 set-up). The deposition was carried out either in the cleanroom (25 °C, 40 % of relative humidity) or using the same device used for the contact angle measurements, giving identical deposited patterns and drop sizes with respect to the volume used.

**Methods.**

*Atomic force microscopy (AFM)* images were recorded with an NT-MDT Solver proequipment. AFM topography was performed in the intermittent contact mode with standard silicon cantilevers. Image analysis was achieved with the free software WSxM.[32]

*Surface Profilometry.* Height profile measurements of the dried patterns were performed using a mechanical profilometer Dektak 150.

*Contact Angle measurements.* Static contact angles were measured under ambient conditions (at 20°C and 40% relative humidity) analyzing the drop profile of sessile drops. A 10 µL droplet of milliQ water was deposited on the sample surface using a Krüss DSA100 apparatus (Germany) equipped with a CCD camera and an image analysis processor. Three droplets were analyzed on different locations on each sample. The reported values are the averages of these measurements for each kind of surface (see Table S1-S5, Figure S4).

*Videos.* Videos of the evaporating processes were captured by a side view CCD camera (DMK 33UX174, Imaging Source, Germany) with a frame rate of 1 Hz. A top view CCD camera in synchronization with the side view one was deployed to monitor the morphology of the evaporation droplets and the corresponding particles, either in the form of films or in the form of rings, left behind on the solid surface. Experimental parameters, such as droplet size, evaporation speed, etc., were extracted from the home-developed package based on matlab. The precision of the target localization is 0.1 pixel with the sub-pixel technique.

*Dynamic light scattering (DLS)* experiments were carried out with a NanoZS apparatus (Malvern Instrument) operating at $\lambda$ = 632.8 nm and P ≤ 4 mW. The scattered intensity was measured at 173 °. Data analysis was carried out by converting the measured intensity autocorrelation function into the scattered electric field autocorrelation function using the Siegert relation. The electric field autocorrelation functions were further analyzed by regularized inverse Laplace transformation using the CONTIN algorithm to yield the distribution of relaxation times ($\tau$).

*Laser Doppler velocimetry.* Electrophoretic mobility ($\mu$) was measured with a NanoZS apparatus (Malvern Inst.). This set-up operates with an electrical field of 25 V.cm$^{-1}$ oscillating successively at 20 Hz and 0.7 Hz to reduce the electroosmosis effect due to the surface charge of the capillary cell. The particles' velocity was measured by laser Doppler velocimetry.

*Scanning electronic microscopy with a field emission gun (SEM-FEG)* images of the deposited nanoparticles were obtained using a SEM-FEG Zeiss Merlin Compact with a resolution of ~ 2 nm at 10 kV. All images are displayed without any post processing.

**Calculation of the DLVO force between a nanoparticle and the substrate.** The electrostatic double layer force ($F_{El.}$) between a spherical particle (radius $R$ and surface potential $\psi_P$) and a flat substrate (surface potential $\psi_S$), separated by a layer of electrolyte aqueous solution (Debye length $\kappa^{-1}$), of thickness $D$, has been estimated using the expression[33]:

$$F_{El.} = \kappa R Z e^{-\kappa D} \qquad (1)$$

where Z (in J.m$^{-1}$) stands for the interaction constant defined for monovalent electrolyte at 25 °C by:

$$Z = 64\pi\varepsilon_0\varepsilon \left(\frac{kT}{e}\right)^2 tanh\left(\frac{e\psi_P}{4kT}\right) tanh\left(\frac{e\psi_S}{4kT}\right) = 9.22 \times 10^{-11} tanh\left(\frac{\psi_P}{103}\right) tanh\left(\frac{\psi_S}{103}\right) \qquad (2)$$

Where $\varepsilon_0\varepsilon$ is the total permittivity of the water, $k$ is the Boltzmann constant, $T$ is the temperature, $e$ the electronic unit charge. All of the suspensions under scrutiny contained a small amount of ionic additives (i.e. Na$^+$ for SiNP$^-$ and AuNP$^-$, Cl$^-$ for SiNP$^+$) at ~ 35 mM in the stock solutions. In our experiments the stock solutions were diluted with milliQ water so that the initial concentration of ions varies together with the initial concentration of particles: $10^{-3} \lesssim$ [ions], mM $\lesssim 11$ for $10^{-2} \lesssim$ [NP], g/L $\lesssim 100$. As a consequence, $\kappa^{-1}$ varied between 300 nm and 3 nm, before drying, in the investigated range of particle concentration assuming that these ions are the only contributors to $\kappa^{-1}$. For the calculation we considered that $\kappa^{-1}(in\ nm) \approx 0.304/\sqrt{[ions]}$ for 1:1 electrolytes at 298 K.[33]

The van der Waals force ($F_{vdW.}$) between a spherical particle (radius $R$) and a flat substrate, separated by a layer of thickness $D$, has been calculated using the expression:

$$F_{vdW.} = -\frac{2HR^3}{3(2R+D)^2 D^2} \qquad (3)$$

With H, the non retarded Hamaker constant for a particle interacting with the substrate across water at room temperature. Negative force implies attraction.

The particle/substrate DLVO force is the algebraic sum of the van der Waals and the electrostatic forces: $F_{DLVO} = F_{vdW.} + F_{El.}$. The evolution of $\kappa^{-1}$ and $F_{DLVO}$, calculated at $t_0$ for D = $\kappa^{-1}$, with the initial particle concentration is plotted in the Figure 1. We underline that the expression (1) used for the calculation of $F_{El.}$ is obtained with the weak overlap approximation which is accurate for surface separations beyond about $\kappa^{-1}$.[33] The choice to work at the limit of $\kappa D = 1$ allows us to correctly estimate the order of magnitude of the force domain covered by the study and to compare it with those already published.

| NP | R nm | $\Psi_{P.}$ mV | Sub | $\Psi_{S.}$ mV | $F_{El.}$ nN | $F_{vdW.}$ nN | $F_{DLVO}$ nN | $\Theta$ ° |
|---|---|---|---|---|---|---|---|---|
| SiNP+ | 37.5 | + 50 | AuSub+ | + 140 | +4.7×10⁻³/+0.47 | -1.4×10⁻⁷/-0.03 | +4.7×10⁻³/+0.44 | 60 |
| | | | AuSub^bare | - 15 | -7.8×10⁻⁴/-0.08 | -1.4×10⁻⁷/-0.03 | -7.8×10⁻⁴/-0.11 | 78 |
| | | | AuSub-- | - 140 | -4.7×10⁻³/-0.47 | -1.4×10⁻⁷/-0.03 | -4.7×10⁻³/-0.50 | 59 |
| SiNP- | 45.0 | - 50 | AuSub+ | + 140 | -5.6×10⁻³/-0.56 | -2.2×10⁻⁷/-0.04 | -5.6×10⁻³/-0.60 | 52 |
| | | | AuSub-- | - 140 | +5.6×10⁻³/0.56 | -2.2×10⁻⁷/-0.04 | +5.6×10⁻³/+0.53 | 53 |
| AuNP- | 15.0 | - 37 | AuSub+ | + 140 | -1.4×10⁻³/-0.14 | -0.7×10⁻⁷/-0.06 | -1.4×10⁻³/-0.20 | 52 |
| | | | AuSub-- | - 140 | +1.4×10⁻³/+0.14 | -0.7×10⁻⁷/-0.06 | +1.4×10⁻³/+0.08 | 53 |

**Table 1.** Nanoparticles (NP) and Substrates (Sub) used in this study with their acronym and main characteristics: particle radius measured by SEM-FEG (R, see figure S1 and S2), particle surface potential ($\Psi_{P.}$) assumed to be equal to the zeta potential knowing that this is an underestimation, substrate surface potential ($\Psi_{S.}$), Hamaker constant Particle/Water/Gold ($H_{in\ water}$), electrostatic double layer force ($F_{El.}$), van der Waals force ($F_{vdW.}$), DLVO force ($F_{DLVO}$) and equilibrium contact angle averaged on the different concentrations ($\Theta$, see Table S1-S5 and Figure S4). We indicate the minimum/maximum forces that have been calculated before drying ($t_0$) for $D = \kappa^{-1}$ at the initial minimum/maximum particle concentrations respectively. The evolution of $\kappa^{-1}$ and $F_{DLVO}$ with the initial particle concentration is plotted in the Figure 1. For comparison, $F_{DLVO}$ considered in reference [9], calculated in the same way, ranged between -0.02 nN and + 0.13 nN. Negative force implies attraction. The energies corresponding to the potentials of interaction and Hamaker constants used for calculations are given in the table S13 and S14 respectively of S.I..

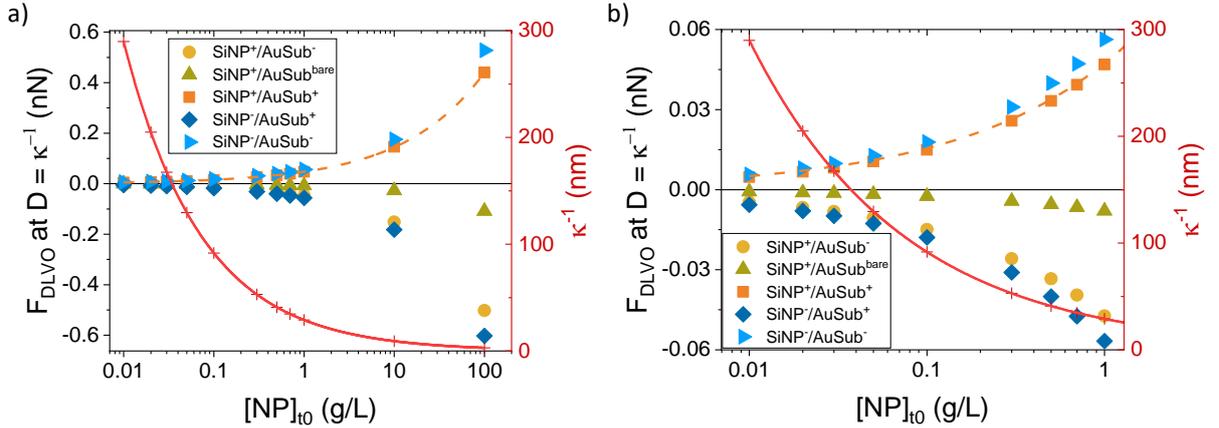

**Figure 1.** a) Evolution of the particle/substrate DLVO force ($F_{DLVO}$) and Debye length ($\kappa^{-1}$, red cross) with the initial particle concentration $[NP]_{t0}$ for different nanoparticles (NP) and substrates (Sub) as indicated in the inset. $F_{DLVO}$ has been calculated before drying at a separation distance $D = \kappa^{-1}$. The continuous red line corresponds to $\kappa^{-1} = 29/\sqrt{[NP]_{t0}}$ and the orange dashed line corresponds to $F_{DLVO} = 0.047\sqrt{[NP]_{t0}}$. b) is an expansion of a) in the diluted domain.

## Results and discussion

**Influence of particles concentration and $F_{DLVO}$ on the dry pattern for $SiNP^+/AuSub^+$.**

We first investigated dry patterns obtained with cationic gold substrates ($AuSub^+$) and cationic silica particles ($SiNP^+$) with initial concentrations varying over 4 decades between 0.01 g/L and 100 g/L (i.e. $4.10^{-4} \leq \Phi_{vol.}(\%) \leq 4$). We placed 0.5 μL of the drops on the gold substrates, and we observed the patterns after complete drying with $t_{drying} \approx 15$ min. In this case study, the particle/substrate interaction is always repulsive with $F_{DLVO}$ ranging between $+4.7 \times 10^{-3}$ and $+0.44$ nN with the initial particle concentration at a separation distance corresponding to the Debye length (Table 1, Figure 1). For clarity, in this part, we will mainly discuss the observed effects with reference to the concentration. The effect of $F_{DLVO}$ will be more specifically discussed in the following section where we will consider the different particle/substrate associations.

Figure 2 shows typical SEM-FEG images of the dry patterns. Overall, we observed a general trend of particle distribution on the substrate as a function of concentration. At low concentrations, particles deposited preferentially in the center of the initial imprint of the drop, forming dot-like patterns, while increasing concentration resulted in a deposit of particles at the periphery, forming a thick ring pattern, and finally a deposition over the entire surface at high concentrations. This concentration-controlled shape transition sequence was reproduced at least 3 times for each concentration under the same conditions (Figure S7-S10) or by varying the injection mode (manual, automated), the volume of the drops (Table S11) and the deposition height from the surface (Table S12).

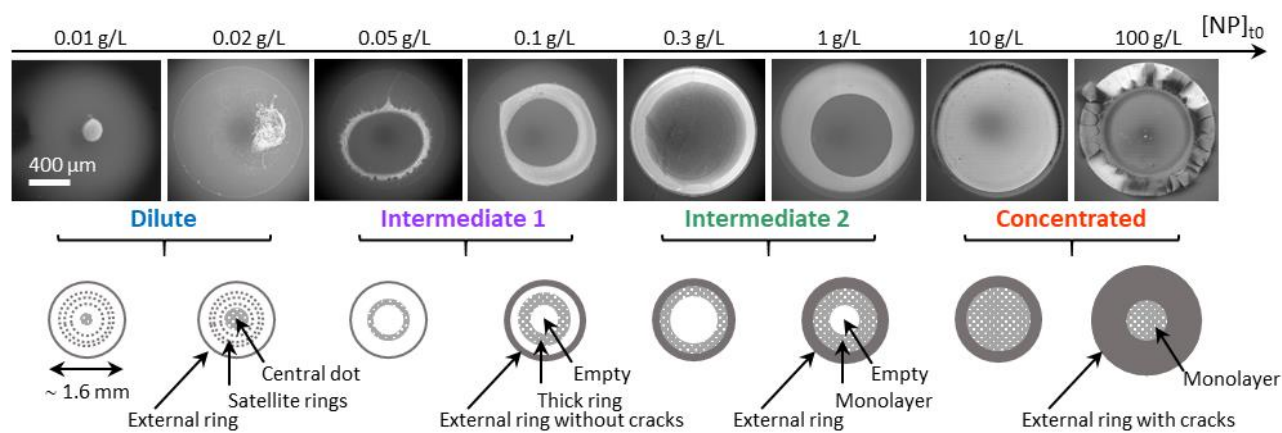

**Figure 2.** Typical SEM-FEG images, and schematic representations, of the dry patterns formed after the evaporation of sessile drops (0.5 μL) of $SiNP^+$ suspensions at different initial concentrations ($[NP]_{t0}$) deposited on gold substrates $AuSub^+$. All images are at the same scale.

We present first our observations in the dilute domain (i.e. $[SiNP^+]_{t0} \leq 0.04 \pm 0.01$ g/L).

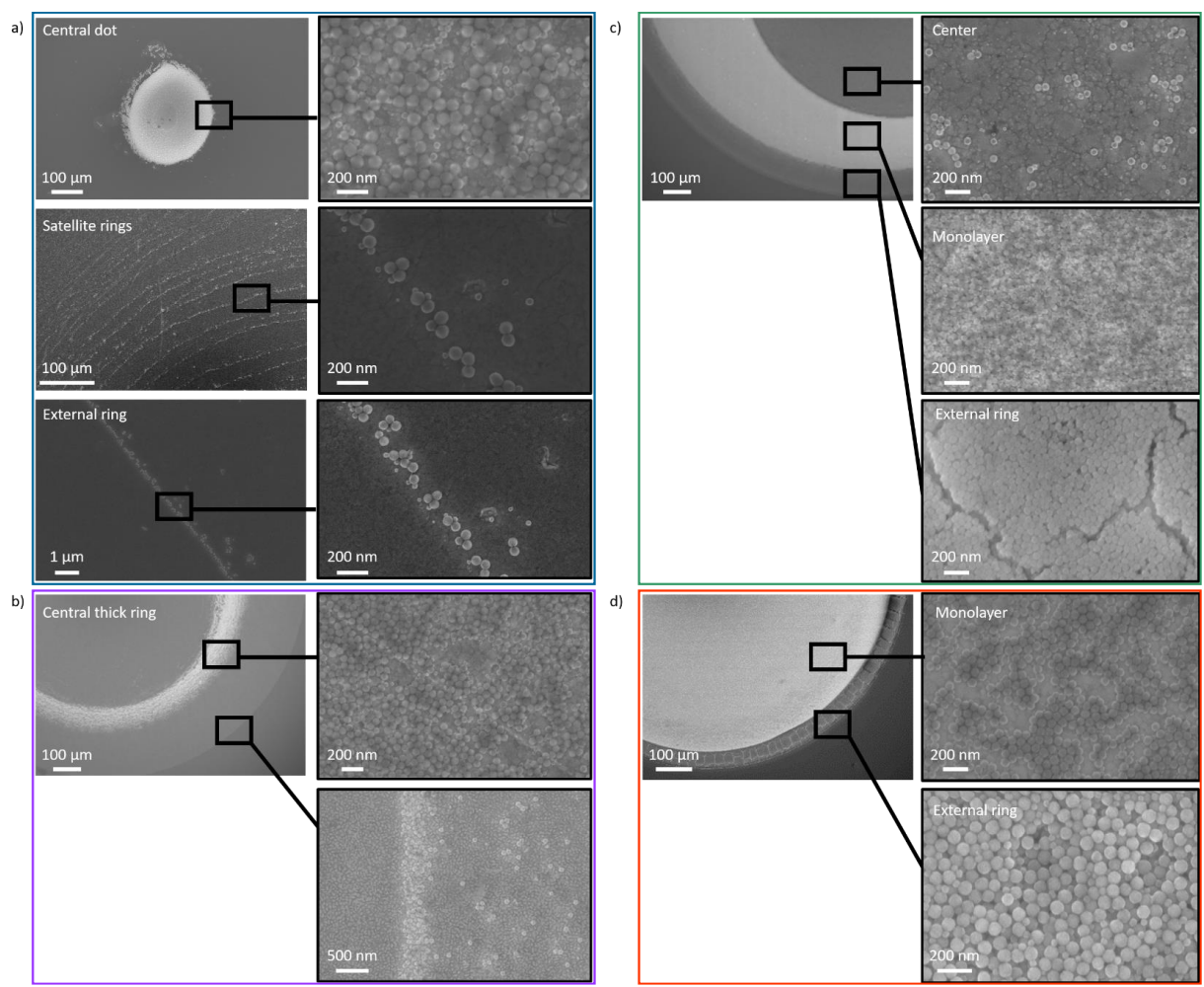

**Figure 3.** Typical SEM images at different magnifications of dry patterns formed after evaporation of 0.5 µL drops of SiNP$^+$ suspensions on AuSub$^+$ in the different concentration domains: (a) dilute ($[SiNP^+]_{t0} < 0.04$ g/L), (b) intermediate 1$^{st}$ part ($0.04 \leq [SiNP^+]_{t0} < 0.3$ g/L), (c) intermediate 2$^{nd}$ part ($0.3 \leq [SiNP^+]_{t0} \leq 1$ g/L) and (d) concentrated ($[SiNP^+]_{t0} \geq 10$ g/L).

Zooming into the dot-like patterns, we observed the presence of satellite rings in between the central dot and the initial edge of the deposited droplet. They are composed of a fine deposit a few particles wide (Figure 3a). According to high magnification images and profilometry measurements (Figure S7, Table S8), the central dot is a continuous domain of particles compacted in a monolayer with holes. This type of organization evokes "spinodal dewetting" surface patterns.[34,35] The shape of the dot is not always circular (i.e. flow patterns could appear, Figure S7 and S9) with a characteristic size increasing with the concentration (Figure S7). For such systems at low particles concentration, with particle/substrate repulsive interactions, ring-like patterns are usually expected. In contrast, more homogeneous deposits are eventually expected in the presence of attractive interactions and/or at high concentration.[1,36–39] One of the possible reasons why these dot-like patterns have never been observed is that the very diluted domain has been little considered so far. A first attempt of explanation is that this type of deposit appears because both the concentration and interactions are very low, which does not allow the formation of a deposit large enough to trap the contact line. On the contrary, the particles must flow with the receding of the contact line until the final stage where most of the particles are collected and form the dot pattern. We will come back to the mechanism of formation of this type of low-concentration deposit in the last part of the article.

Now, we switch to the intermediate regime with the initial concentration ranging from 0.05 g/L to 0.10 g/L (i.e. intermediate 1st part with $\Phi_{vol.}$ ~ $2.10^{-3}$ - $4.10^{-3}$ %). A transition of pattern morphology towards thick ring patterns occurred. The patterns observed in this domain present a peculiar structure taking the form of a thick ring whose outer diameter is much smaller than the diameter of the initial deposited droplet (Figure 3b). The disk situated at the center of the thick ring is free of particles, whereas fine satellite rings could be detected beyond the outer edge. It is to be noted that this particular structure was not systematically observed, although we recorded several times (see Figure S8-9), as it occurs for a narrow range of particles concentration. When the concentration is increased beyond the threshold, the ring widens until it reaches the initial edge of the initially placed drop for $[SiNP^+]_{t0}$ ~ 0.3 g/L (i.e. $\Phi_{vol.}$ ~ $1.2 \times 10^{-2}$ %). Interestingly, we point out that the number of particles required to form a monolayer of particles, with a packing density of 0.8, on the entire surface initially covered by the drop corresponds to an initial particle concentration of 0.32 g/L. Beyond this concentration, the central disk, which until then had remained empty of particles, while keeping almost the same surface, is progressively covered with a monolayer of particles. The surface is completely covered from $[SiNP^+]_{t0}$ ~ 10 g/L (i.e. $\Phi_{vol.}$ ≈ 0.4 %). To sum up, an external ring of particles with an outer diameter corresponding to the initial diameter of the deposited drop was observed regardless of the particle concentration. It is always separated from the rest of the deposit. At high magnification, one observes that the particles packing decreases going from the outside to the inside of the ring.[40] As expected from previous studies,[36,41] the width (w, see Figure 4c) of this external ring increases as power law with the particle concentration (Figure 4a) as follows: $w/r \propto \phi_{vol.}^{0.55}$ with r, the initial radius of the deposited drop. Although the structure of the ring is heterogeneous (i.e. variation of the particle packing) in the radial direction, the obtained power law exponent (β) is in good agreement with the theoretical prediction from Popov[42], based on the conservation of the mass of droplet liquid and particles during the evaporation. This model predicts a square root dependence of w with the initial volume fraction, as $w/r \propto$ $\left(\phi/\rho\right)^{0.5}$ with ρ the particle packing fraction.[41] We underline that the width of the ring does not respect the power law evolution at the beginning of the intermediate domain and sometimes also in the dilute domain. Qualitatively, the power law evolution indicates that NPs accumulate in greater quantities near the triple contact line during the first 80% of the evaporation time when the initial particle concentration is increased.

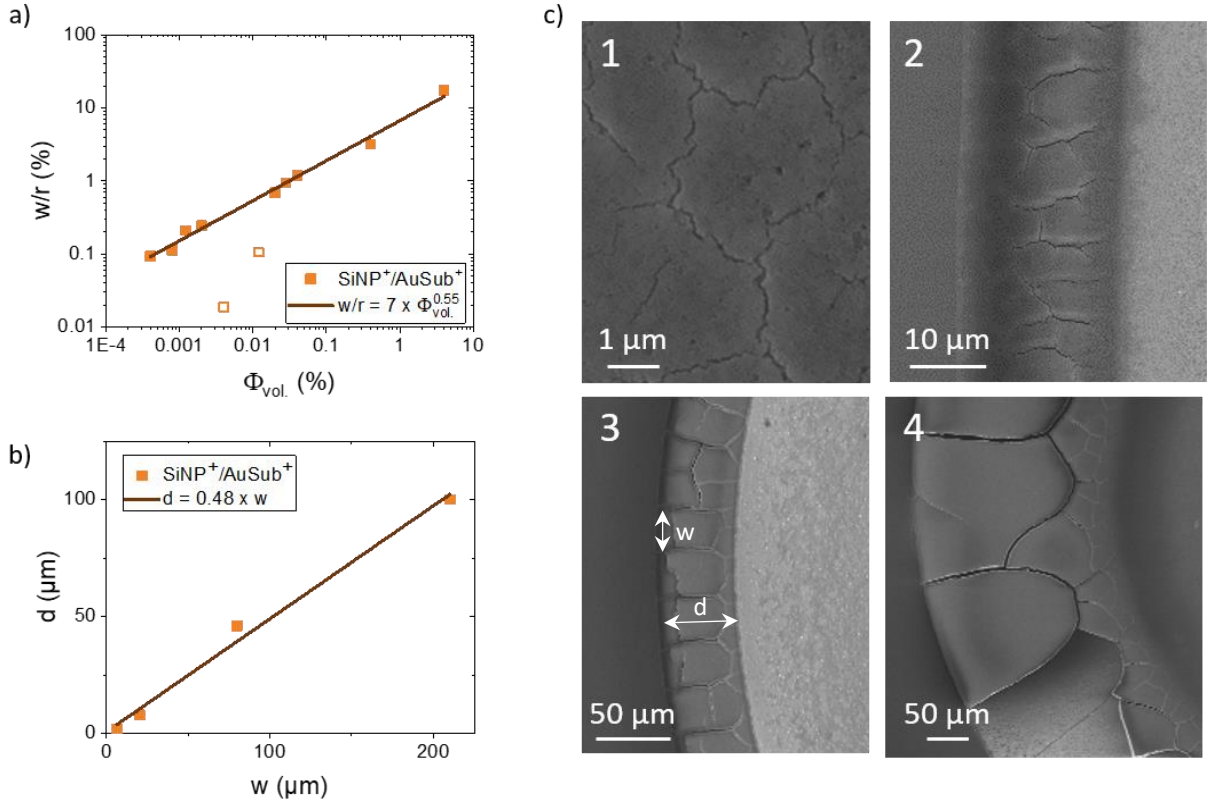

**Figure 4.** External ring characterizations. a) Dimensionless ring width [w/r] as a function of the particle concentration. Open symbols correspond to samples of the intermediate domain that were not taken into account for the fit. b) Crack spacing (d) plotted as a function of the ring width (w). c) SEM-FEG images of the ring top view at different particles concentrations: (1) 0.5 g/L, (2) 1 g/L, (3) 10 g/L, (4) 100 g/L.

When the height of the external ring is large enough, for $[SiNP^+]_{t0} \geq 0.5$ g/L, we observe cracks at fairly regular intervals with a radial orientation (Figure 4c). This radial orientation agrees with our low charge screening condition.[43] We underline the presence of secondary ortho-radial cracks connected to the main radial crack pattern at the two lowest concentrations for which the system must relax stress more frequently. The crack spacing measured on the outside circumference line (noted d, see Figure 4c) increases with the particle concentration and, as a consequence with w, in a linear manner (Figure 4b). This result agrees with previous studies[44,45] assuming that the height and the width scale in the same manner with concentration.[41,42]

Overall, this first part of the study allowed us to identify original patterns (i.e. dot-like patterns and a central thick ring with fine satellite rings) in the dilute domain and to retrieve several known results at higher concentrations, which assures us of the method robustness.

**Influence of NP/Sub DLVO Force in the Different Concentration Domains.**

We continued by examining more specifically the influence of NP/Sub DLVO force on the dry pattern morphology in the different concentration regimes identified in the previous section.

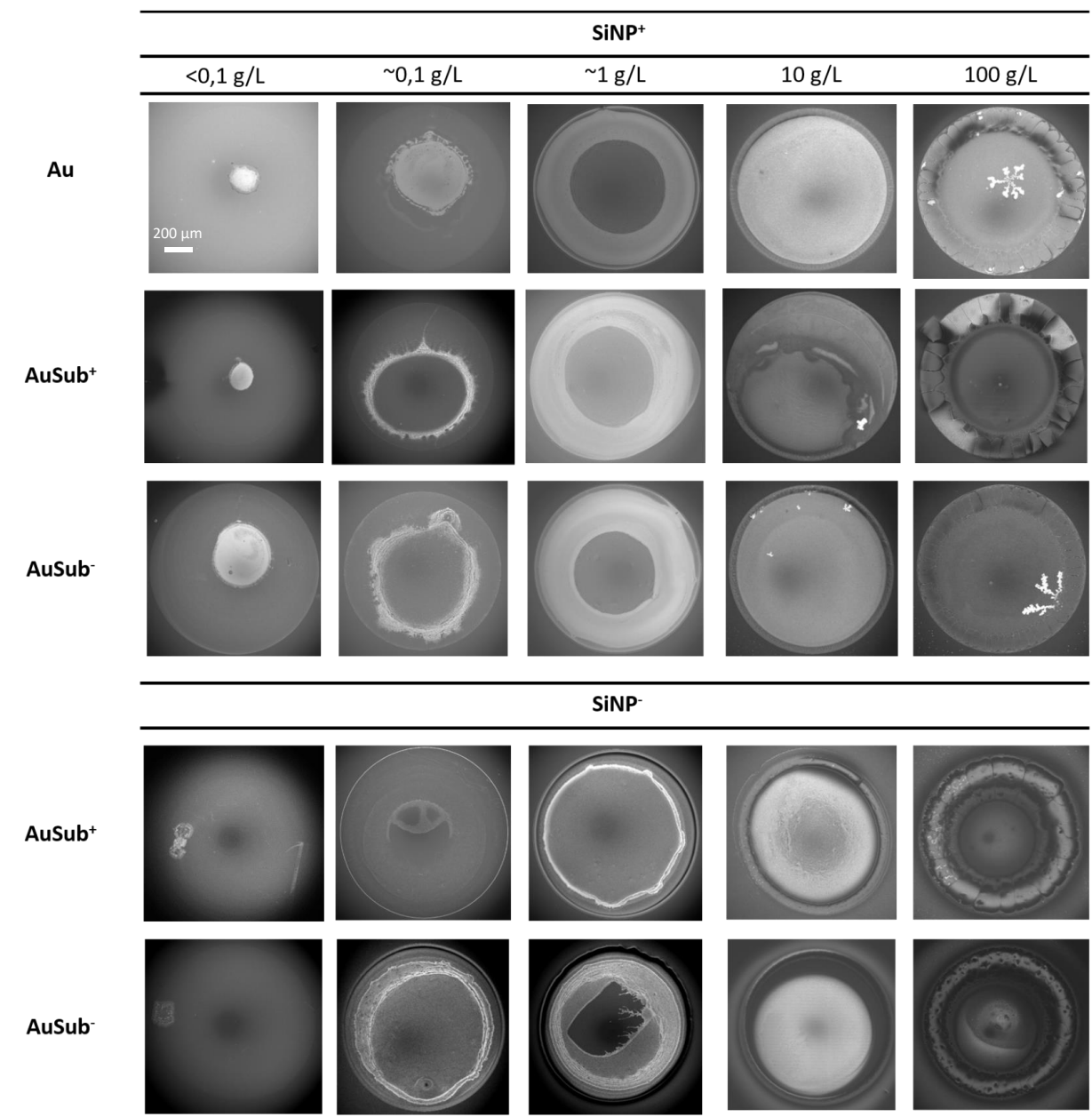

**Figure 5.** Typical SEM images of dry patterns formed after the evaporation of sessile drops (0.5 μL) of SiNP$^+$ and SiNP$^-$ solutions at different concentrations on bare Au substrate (i.e. AuSub$^{bare}$), AuSub$^+$ and AuSub$^-$ as indicated in the figure. All images are at the same scale.

This work was carried out considering three types of particles and three types of substrates (see Table 1) with the same experimental method. The equilibrium contact angle varies little with particles concentration and substrate (i.e. $55 \leq \Theta (°) \leq 65$), except for bare Au substrate (i.e. $70 \leq \Theta (°) \leq 90$). Figure 5 shows typical SEM-FEG images of the dry patterns for $SiNP^+$ and $SiNP^-$ solutions of various concentrations on the different substrates. The evolution of the dry pattern morphology as a function of the particles concentration and DLVO force is summarized in the Figure 6, in the form of a state diagram. Overall, we observed the same general trend of particle distribution on the substrate as a function of concentration whatever the sign and magnitude of the NP/Sub DLVO force. In all cases, we observed the same general trend of particle distribution on the substrate as a function of concentration as described in the previous section for $SiNP^+/AuSub^+$. Finally, the concentration boundaries between the different domains vary a little from one system to another but with no obvious correlation with DLVO force.

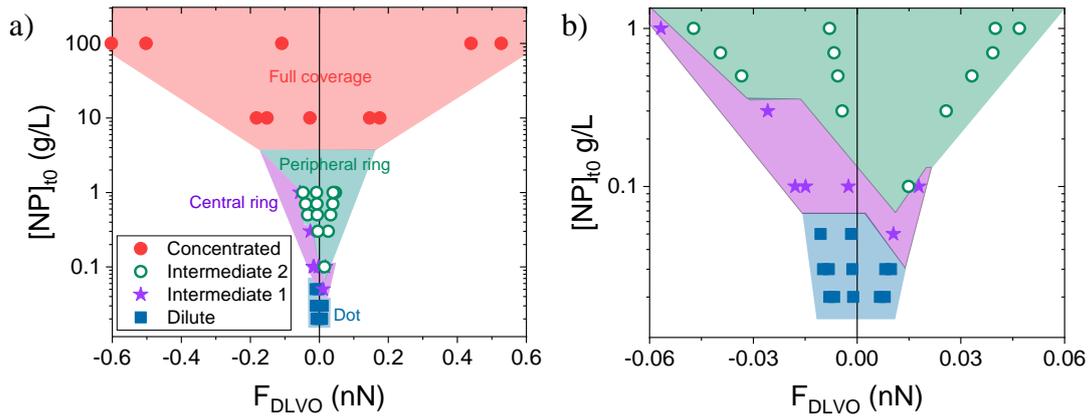

**Figure 6.** a) Schematic pattern diagram summarizing the dry pattern morphologies as a function of the initial particles concentration $[NP]_{t0}$ and of the particle/substrate DLVO force ($F_{DLVO}$) calculated before drying at a separation distance $D = \kappa^{-1}$ as already shown in the Figure 1. b) is an expansion of a) in the dilute domain.

These results show that, contrarily to what has been proposed in several previous studies, the NP/Sub DLVO force does not significantly influence the general morphology of the dry deposits in the different concentration areas, at least in the domain of forces under scrutiny in the absence of long distance attractive NP/NP interactions. Furthermore, we quantified the relative contribution of convection and diffusion to particle transport by estimating the Péclet number which is of the order of *Pe* ~790 for $SiNP^+$ and $SiNP^-$ showing that NPs transport in evaporation droplet is driven by the convection flow that brings the NPs to the contact line. This strong convection flow should overwhelm the DLVO forces and dominates the pattern formation. Moreover, we observed that the addition of salt, as low as 10 mM (Figure S10 and S13), induced the formation of uniform deposits mixing particles and salt crystals instead of ring patterns whatever the system. This shows that the pattern morphology can be easily modified by a screening of NP/NP interactions. A study taking into account the contribution of this dimension would be relevant in the future given the screening conditions encountered in biofluids or just in buffered media.

We then conducted a more quantitative study on certain characteristic lengths of the structures formed. We first considered the relation between the width (w) of the thick ring, observed in the intermediate domain of concentration, and the particles concentration (Figure 7a). In all cases, w increases as power law with an exponent close to the expected square root dependence. From this figure it is also clear that the power laws are identical for a given type of particle whatever the substrate and thus, whatever the NP/Sub DLVO force. Interestingly, slight variations were observed between the different set of particles suggesting that the power law could be influenced by the particle size as illustrated in Figure 7a where data from Deegan[41] and Brutin[36] were added for comparison. It may reflect the influence of NP/NP interaction and/or polydispersity on the particle packing (i.e. heterogeneity between the layers). These effects appear credible in the light of past studies.[40,46]

We then studied the cracks for the different combinations of particles and substrates with the exception of the gold particles for which we were unable to probe regimes sufficiently concentrated in particles to have outer rings of sufficient thickness for the cracks to appear. At high concentrations, typically for $[SiNP]_{t0} \geq 0.5$ g/L, the external rings have cracks similar to those described in the previous section whatever the combination of particle and substrate (Figure S14). We did not observe a significant difference in the shape of the cracks from one system to the other with predominantly radial patterns and an increasing proportion of secondary orthoradial patterns moving towards low concentrations (i.e. low thicknesses). In all cases, a linear relation exists between the crack spacing, d, and the width, w, of the external ring (Figure 7b). The relations are clearly different for the same type of particle depending on the substrate. Nevertheless, this effect does not seem to be correlated to the NP/Sub DLVO force but more to the nature of the substrate. Thus, films deposited on the anionic substrates must relax stress less frequently than films deposited on the cationic substrates (i.e. lower d for same w for films on $AuSub^-$ than for films on $AuSub^+$). We do not have a clear explanation for this result since the equilibrium contact angles are globally the same on these two surfaces.

Finally, we also considered the drying of aqueous drops containing cationic or anionic gold nanoparticles (see Table 1) on different substrates. In general, the contrast of the images (i.e. gold on gold) and the small size of the particles complicate the observations as shown in Figure S15. Nevertheless, it appears that the succession of the different deposit morphologies with particle concentration is roughly the same as for silica. A particularity of gold particles is that, in diluted or even semi-diluted regime, they are deposited in the form of small clusters of nanoparticles forming monolayer islands and rarely in the form of a single dot. This may reflect an eventual contribution of the van der Walls force which is greater for gold particles compared to silica particles (see Table 1).

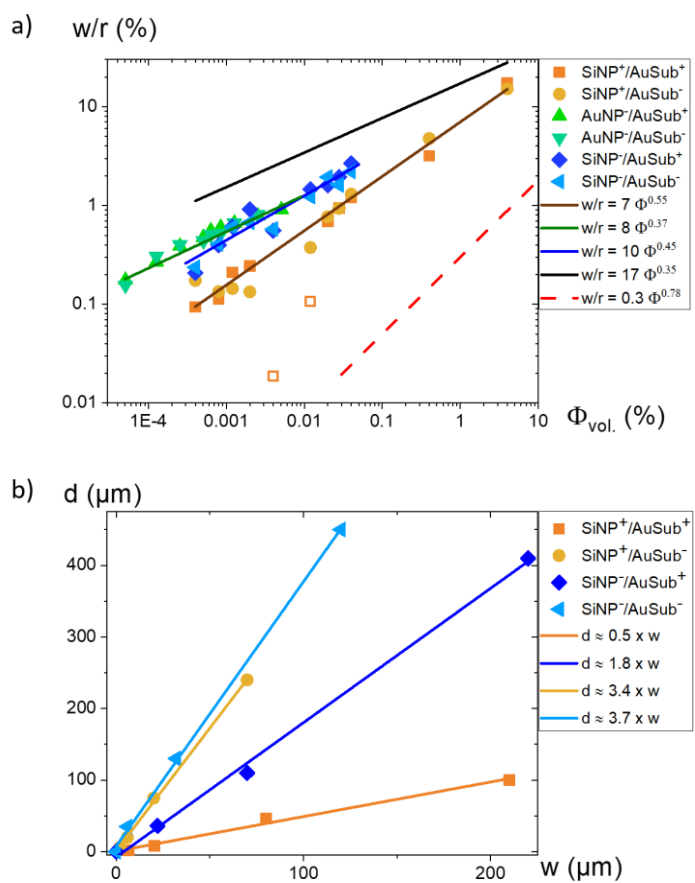

**Figure 7.** External ring characterizations. a) Dimensionless ring width [w/r] as a function of the particle concentration for different NP/Sub combinations. The continuous black line and the dashed red line are power law evolutions obtained by Brutin,[36] with 24 nm carboxylate-terminated polystyrene particles on Nuflon substrate, and Deegan,[41] with 100 nm sulfate-terminated polystyrene particles on mica substrate, respectively. b) Crack spacing (d) plotted as a function of the ring width for different NP/Sub combinations.

Overall, the most striking feature of this second part is that, contrarily to expectations from previous studies dedicated to DLVO interaction, the different shapes of deposits obtained by modifying the particle concentration are the same in the different regimes of concentration regardless of particle/substrate DLVO interaction in the studied range of forces and particle concentrations. Slight effects could be detected on characteristic lengths (i.e. w/r, d) from one system to another but without a clear relation with DLVO force. These results are consistent with the observation of Anyfantakis et al.[47] that the morphology of the dry deposit is mainly determined by interactions between particles in the mass, whereas, in their case, they studied the influence of particle wettability and concentration on the shape of the deposit.

**Mechanisms of Patterns Formation.**

In this last part, we focus on the mechanism of pattern formation at low and high particle concentrations. At low concentrations, we observed an unusual combination of dot-like and multi-rings patterns. In their review, Parsa et al. underlined that dot-like patterns were typically observed on hydrophobic substrates resulting from an evaporation process at constant contact angle with an important contribution of the Marangoni flow that should be oriented radially inward along the substrate and, radially outward along the air/liquid interface.[48] Nevertheless, similar patterns, by other names (i.e. 'dome', 'coffee eyes', "central bump") were observed on hydrophilic substrates with nanoparticles showing an hydrophobic character (and only at high particle concentration)[47] and with hydrophilic particles deposited on hot substrates[49] or mixed with micrometric particles[37]. In all these cases, the observations made by optical microscopy did not bring to light the coexistence of central deposit and concentric rings. On the other hand, multi-rings pattern is the dominant morphology for charged nanoparticles deposited on hydrophilic substrate at low concentration in usual conditions of drying (i.e. no heating). These patterns occur when colloidal droplets evaporate with a stick-slip motion of the contact line. One of the few examples of deposits combining a central cluster and concentric rings has been given with λ-DNA fluorescent labeled molecules.[50] Observations with a confocal laser scanning microscope during evaporation revealed a pinning-depinning cycle in which the position of a new contact line is predetermined by the prior formation of a 'precipitated' DNA ring. However, contrarily to our case, these patterns were only observed in the concentrated domain of DNA concentration (i.e. C ≥ $4.10^{-2}$ g/L at 25 °C).

To clarify the picture of the "dot-like" pattern formation and shed light on a possible dependency of the drying dynamics with $F_{DLVO}$, we recorded the side and top profiles of evaporating drops ($V_0$ = 3 μL). We first studied the drying dynamic for samples in the different ranges of particles concentrations and different particles/substrates (Figure 8a). Although the number of samples considered was limited, it appeared that the dynamics were little dependent on the concentration and $F_{DLVO}$ for the first 90 % of the drying time.

As shown in Figure 8b and 8c, we then extracted the evolution of both contact angle and drop radius as a function of time in the dilute regime ([NP] = 0.03 g/L) and also in the concentrated one ([NP] = 10 g/L) for comparison. This was done for SiNP$^+$ suspension evaporating on AuSub$^+$. At high particle concentration (Figure 8b), the contact line is pinned, so that the wetted contact area between the droplet and substrate remain constant, and the contact angle decreases as an exponential with time. The drop becomes turbid a few seconds before the end of the drying process. This change in appearance suggests the occurrence of phase separation and gelling. The first radial cracks form at about the same time as the drop center still has a liquid character in accordance with past studies.[43]

In the dilute regime, the same evolutions were observed for 90 % of the time and at 100 seconds from complete drying, a stick-slip dynamic takes place as illustrated in the insets of figure 8c. During this period, the contact line moves step by step and the contact angle oscillates in a synchrone manner with a continuous decrease during each pinning step. The drop becomes turbid only a few seconds before the end of this stepwise process.

This part of the study suggests that the morphologies observed at low concentrations are the result of a somehow classical process where flow should convect particles to the drop edge and deposit them near the contact line. As the drop volume decreases, the contact angle decreases inducing an inward, unbalanced, depinning force on the contact line. When the contact angle approaches a critical value ($\Theta_{Rec.} \sim$ 5-7 °), the contact line slips toward the center until next pinning. The repetition of this cycle generates the formation of concentric rings. Finally, when a certain concentration of particles and ions is reached, the particles aggregate to form the central deposit.

Observations on the side and on the top of the drop suggest a phase separation in the whole drop volume. Nevertheless, the spatial and time resolution of our observations does not allow us to rule out a phase separation initiated in hydrodynamic stagnation domains in the final stage of the evaporation process as proposed in other cases leading to dot like patterns.[49–51] However, here, we think we can rule out the contribution of an inward thermal Marangoni flow in the formation of the central deposit considering the formation of multi-rings and the very high value of the relative thermal conductivity between gold substrate and water (i.e. k$_R \sim$ 200).

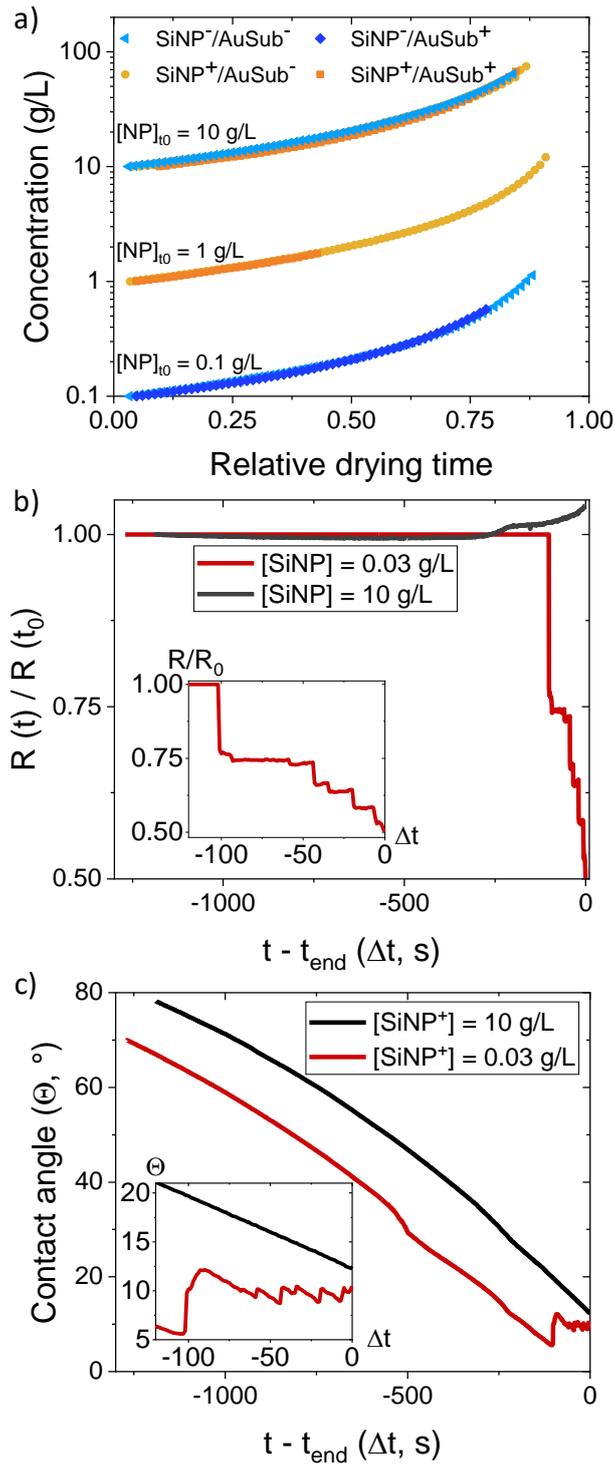

**Figure 8.** (a) Particles concentration in a drying drop (V ≈ 3 μL) vs. the relative drying time for different particles and substrates. The concentration has been calculated from the time variation of the drop volume knowing the concentration at a relative drying time of zero. (b) Time variation of the drop radius R to the initial radius R($t_0$) of the sessile drop deposited on AuSub$^+$. The increase of the black curve in the last stage can be ignored because of our imaging technique limitation, where it should be a constant. (c) Time variation of the contact angle for the same drops as those described in (b). Measurements were done for R($t_0$) ≈ 2.5 mm, at 25 °C, with a relative humidity of ~ 40%.

## Conclusion

The main purpose of this work was to show the dependence of the shape of solid deposits obtained by free drying of sessile drops on the particle concentration and the Derjaguin–Landau–Verwey–Overbeek (DLVO) particle/substrate interaction. In this issue, we considered model substrates and particles of different nature, surface chemistry and size (in the case of particles), whereas, long-distance repulsive interactions between particles were maintained for most of the drying time. For each set of particles and substrate, we varied the particles concentration over unprecedented 4 decades to identify the effect of interactions in the different concentration domains. Three regimes of the drying morphology are revealed: dilute/intermediate/concentrate regime. The details of depositions corresponding to each regime are analyzed with SEM-FEG imaging. Size of the periphery outer ring, including its width and the crack length, is quantitatively reported and compared with previous investigations. Moreover, the dynamic deposition process is monitored with the top view and side view to gain insight into the formation of drying patterns.

Our main result is that the different shapes of deposits obtained by modifying the particles concentration are the same regardless of particle/substrate interaction in the studied range of DLVO forces and particle concentrations.

The second result is that, contrary to expectations, the dominant shapes of deposits at low particles concentration is dot-like patterns for all the studied systems. Zooming into the dot, it appears that they are a continuous domain of particles compacted in a monolayer. We show that this structure results from a stick-slip evaporation process.

We believe that our results and methodology pave the way toward a better understanding of complex fluids drying and in particular the drying of biofluids where different interactions are intertwined, complicating the analysis of the results and the establishment of a diagnosis. Our work also offers new possibilities for a fine modelling from a theoretical point of view and for nanostructuring by working at very low particles concentrations.

## Supporting Information

Characterizations of the nanoparticles and substrates. Contact Angle measurements for the different systems at different particle concentrations. SEM-FEG images of dry patterns for the different systems at different particle concentrations. SEM-FEG images of dry patterns for the different systems at different particle concentrations and salt concentration. SEM-FEG images of dry patterns on lithographed rough surfaces. Profilometry performed on dry patterns for the different systems at different concentrations. Side-view imaging of the evaporation of a 3 μL drop containing 0.03 mg/mL of SiNP$^+$ on AuSub$^+$ corresponding to Figure 8 b-c. Side-view imaging of the evaporation of a 3 μL drop containing 10 mg/mL of SiNP$^+$ on AuSub$^+$ corresponding to Figure 8 b-c.

## Acknowledgments

ANR (Agence Nationale de la Recherche) and CGI (Commissariat à l'Investissement d'Avenir) are acknowledged for their financial support through Labex SEAM (Science and Engineering for Advanced Materials and devices), ANR-10-LABX-096 and ANR-18-IDEX-0001". FC thanks the ANR Coligomere-18-CE06-0006 for funding. We are indebted to T. Sonoda and Nissan Chemicals who kindly provided us the silica nanoparticles. We acknowledge Stephan Suffit (laboratory "Matériaux et Phénomènes Quantiques") for his kind assistance for SEM-FEG, profilometry and surface lithography in the MPQ clean room. We acknowledge Vincent Humblot ("Franche-Comté Electronique Mécanique Thermique et Optique – Sciences et Technologies" institute) for his kind assistance for wetting angle measurements.

## Author Contributions

N.B. and M.Z. contributed equally to this work.

## References

(1) Parsa, M.; Harmand, S.; Sefiane, K. Mechanisms of Pattern Formation from Dried Sessile Drops. *Advances in Colloid and Interface Science* **2018**, *254*, 22–47. https://doi.org/10.1016/j.cis.2018.03.007.
(2) DROPLET. *Droplet Wetting and Evaporation: From Pure to Complex Fluids*; Brutin, D., Ed.; Elsevier/AP, Academic Press: Amsterdam Boston Heidelberg, 2015.
(3) Larson, R. G. Transport and Deposition Patterns in Drying Sessile Droplets. *AIChE Journal* **2014**, *60* (5), 1538–1571. https://doi.org/10.1002/aic.14338.
(4) Mampallil, D.; Eral, H. B. A Review on Suppression and Utilization of the Coffee-Ring Effect. *Advances in Colloid and Interface Science* **2018**, *252*, 38–54. https://doi.org/10.1016/j.cis.2017.12.008.
(5) Zang, D.; Tarafdar, S.; Tarasevich, Y. Y.; Dutta Choudhury, M.; Dutta, T. Evaporation of a Droplet: From Physics to Applications. *Physics Reports* **2019**, *804*, 1–56. https://doi.org/10.1016/j.physrep.2019.01.008.
(6) Kuang, M.; Wang, L.; Song, Y. Controllable Printing Droplets for High-Resolution Patterns. *Advanced Materials* **2014**, *26* (40), 6950–6958. https://doi.org/10.1002/adma.201305416.
(7) Perelaer, J.; Schubert, U. S. Novel Approaches for Low Temperature Sintering of Inkjet-Printed Inorganic Nanoparticles for Roll-to-Roll (R2R) Applications. *J. Mater. Res.* **2013**, *28* (4), 564–573. https://doi.org/10.1557/jmr.2012.419.
(8) Bridonneau, N.; Mattana, G.; Noel, V.; Zrig, S.; Carn, F. Morphological Control of Linear Particle Deposits from the Drying of Inkjet-Printed Rivulets. *The Journal of Physical Chemistry Letters* **2020**. https://doi.org/10.1021/acs.jpclett.0c01244.
(9) Kuncicky, D. M.; Velev, O. D. Surface-Guided Templating of Particle Assemblies Inside Drying Sessile Droplets †. *Langmuir* **2008**, *24* (4), 1371–1380. https://doi.org/10.1021/la702129b.

(10) Anyfantakis, M.; Baigl, D. Manipulating the Coffee-Ring Effect: Interactions at Work. *ChemPhysChem* **2015**, *16* (13), 2726–2734. https://doi.org/10.1002/cphc.201500410.

(11) Bhardwaj, R.; Fang, X.; Somasundaran, P.; Attinger, D. Self-Assembly of Colloidal Particles from Evaporating Droplets: Role of DLVO Interactions and Proposition of a Phase Diagram. *Langmuir* **2010**, *26* (11), 7833–7842. https://doi.org/10.1021/la9047227.

(12) Dugyala, V. R.; Basavaraj, M. G. Control over Coffee-Ring Formation in Evaporating Liquid Drops Containing Ellipsoids. *Langmuir* **2014**, *30* (29), 8680–8686. https://doi.org/10.1021/la500803h.

(13) Moraila-Martínez, C. L.; Cabrerizo-Vílchez, M. A.; Rodríguez-Valverde, M. A. The Role of the Electrostatic Double Layer Interactions in the Formation of Nanoparticle Ring-like Deposits at Driven Receding Contact Lines. *Soft Matter* **2013**, *9* (5), 1664–1673. https://doi.org/10.1039/C2SM27040D.

(14) Homede, E.; Zigelman, A.; Abezgauz, L.; Manor, O. Signatures of van Der Waals and Electrostatic Forces in the Deposition of Nanoparticle Assemblies. *The Journal of Physical Chemistry Letters* **2018**, *9* (18), 5226–5232. https://doi.org/10.1021/acs.jpclett.8b02052.

(15) Homede, E.; Manor, O. Deposition of Nanoparticles from a Volatile Carrier Liquid. *Journal of Colloid and Interface Science* **2020**, *562*, 102–111. https://doi.org/10.1016/j.jcis.2019.11.062.

(16) Wang, W.; Yin, Y.; Tan, Z.; Liu, J. Coffee-Ring Effect-Based Simultaneous SERS Substrate Fabrication and Analyte Enrichment for Trace Analysis. *Nanoscale* **2014**, *6* (16), 9588. https://doi.org/10.1039/C4NR03198A.

(17) Xu, J.; Du, J.; Jing, C.; Zhang, Y.; Cui, J. Facile Detection of Polycyclic Aromatic Hydrocarbons by a Surface-Enhanced Raman Scattering Sensor Based on the Au Coffee Ring Effect. *ACS Applied Materials & Interfaces* **2014**, *6* (9), 6891–6897. https://doi.org/10.1021/am500705a.

(18) Sefiane, K. On the Formation of Regular Patterns from Drying Droplets and Their Potential Use for Bio-Medical Applications. *Journal of Bionic Engineering* **2010**, *7* (S4), S82–S93. https://doi.org/10.1016/S1672-6529(09)60221-3.

(19) Yakhno, T. A.; Sanin, A. A.; Ilyazov, R. G.; Vildanova, G. V.; Khamzin, R. A.; Astascheva, N. P.; Markovsky, M. G.; Bashirov, V. D.; Yakhno, V. G. Drying Drop Technology as a Possible Tool for Detection Leukemia and Tuberculosis in Cattle. *Journal of Biomedical Science and Engineering* **2015**, *08* (01), 1–23. https://doi.org/10.4236/jbise.2015.81001.

(20) Devineau, S.; Anyfantakis, M.; Marichal, L.; Kiger, L.; Morel, M.; Rudiuk, S.; Baigl, D. Protein Adsorption and Reorganization on Nanoparticles Probed by the Coffee-Ring Effect: Application to Single Point Mutation Detection. *Journal of the American Chemical Society* **2016**, *138* (36), 11623–11632. https://doi.org/10.1021/jacs.6b04833.

(21) Trantum, J. R.; Wright, D. W.; Haselton, F. R. Biomarker-Mediated Disruption of Coffee-Ring Formation as a Low Resource Diagnostic Indicator. *Langmuir* **2012**, *28* (4), 2187–2193. https://doi.org/10.1021/la203903a.

(22) Yan, Q.; Gao, L.; Sharma, V.; Chiang, Y.-M.; Wong, C. C. Particle and Substrate Charge Effects on Colloidal Self-Assembly in a Sessile Drop. *Langmuir* **2008**, *24* (20), 11518–11522. https://doi.org/10.1021/la802159t.

(23) Denkov, N.; Velev, O.; Kralchevski, P.; Ivanov, I.; Yoshimura, H.; Nagayama, K. Mechanism of Formation of Two-Dimensional Crystals from Latex Particles on Substrates. *Langmuir* **1992**, *8* (12), 3183–3190. https://doi.org/10.1021/la00048a054.

(24) Zigelman, A.; Manor, O. The Deposition of Colloidal Particles from a Sessile Drop of a Volatile Suspension Subject to Particle Adsorption and Coagulation. *Journal of Colloid and Interface Science* **2018**, *509*, 195–208. https://doi.org/10.1016/j.jcis.2017.08.088.

(25) Lee, H. H.; Fu, S. C.; Tso, C. Y.; Chao, C. Y. H. Study of Residue Patterns of Aqueous Nanofluid Droplets with Different Particle Sizes and Concentrations on Different Substrates. *International Journal of Heat and Mass Transfer* **2017**, *105*, 230–236. https://doi.org/10.1016/j.ijheatmasstransfer.2016.09.093.

(26) Bastús, N. G.; Comenge, J.; Puntes, V. Kinetically Controlled Seeded Growth Synthesis of Citrate-Stabilized Gold Nanoparticles of up to 200 Nm: Size Focusing versus Ostwald Ripening. *Langmuir* **2011**, *27* (17), 11098–11105. https://doi.org/10.1021/la201938u.

(27) Shi, L.; Buhler, E.; Boue, F.; Carn, F. How Does the Size of Gold Nanoparticles Depend on Citrate to Gold Ratio in Turkevich Synthesis? Final Answer to a Debated Question. *Journal of Colloid and Interface Science* **2017**, *492*, 191–198. https://doi.org/10.1016/j.jcis.2016.10.065.

(28) Lin, S.-Y.; Tsai, Y.-T.; Chen, C.-C.; Lin, C.-M.; Chen, C. Two-Step Functionalization of Neutral and Positively Charged Thiols onto Citrate-Stabilized Au Nanoparticles. *The Journal of Physical Chemistry B* **2004**, *108* (7), 2134–2139. https://doi.org/10.1021/jp036310w.

(29) Mercier, D.; Boujday, S.; Annabi, C.; Villanneau, R.; Pradier, C.-M.; Proust, A. Bifunctional Polyoxometalates for Planar Gold Surface Nanostructuration and Protein Immobilization. *The Journal of Physical Chemistry C* **2012**, *116* (24), 13217–13224. https://doi.org/10.1021/jp3031623.

(30) Ben Haddada, M.; Huebner, M.; Casale, S.; Knopp, D.; Niessner, R.; Salmain, M.; Boujday, S. Gold Nanoparticles Assembly on Silicon and Gold Surfaces: Mechanism, Stability, and Efficiency in Diclofenac Biosensing. *The Journal of Physical Chemistry C* **2016**, *120* (51), 29302–29311. https://doi.org/10.1021/acs.jpcc.6b10322.

(31) Valotteau, C.; Calers, C.; Casale, S.; Berton, J.; Stevens, C. V.; Babonneau, F.; Pradier, C.-M.; Humblot, V.; Baccile, N. Biocidal Properties of a Glycosylated Surface: Sophorolipids on Au(111). *ACS Applied Materials & Interfaces* **2015**, *7* (32), 18086–18095. https://doi.org/10.1021/acsami.5b05090.

(32) Horcas, I.; Fernández, R.; Gómez-Rodríguez, J. M.; Colchero, J.; Gómez-Herrero, J.; Baro, A. M. WSXM : A Software for Scanning Probe Microscopy and a Tool for Nanotechnology. *Review of Scientific Instruments* **2007**, *78* (1), 013705. https://doi.org/10.1063/1.2432410.

(33) Israelachvili, J. N. *Intermolecular and Surface Forces*, Third edition.; Elsevier, Academic Press: Amsterdam, 2011.

(34) Xie, R.; Karim, A.; Douglas, J. F.; Han, C. C.; Weiss, R. A. Spinodal Dewetting of Thin Polymer Films. *Physical Review Letters* **1998**, *81* (6), 1251–1254. https://doi.org/10.1103/PhysRevLett.81.1251.

(35) Wyart, F. B.; Daillant, J. Drying of Solids Wetted by Thin Liquid Films. *Canadian Journal of Physics* **1990**, *68* (9), 1084–1088. https://doi.org/10.1139/p90-151.

(36) Brutin, D. Influence of Relative Humidity and Nano-Particle Concentration on Pattern Formation and Evaporation Rate of Pinned Drying Drops of Nanofluids. *Colloids and Surfaces A: Physicochemical and Engineering Aspects* **2013**, *429*, 112–120. https://doi.org/10.1016/j.colsurfa.2013.03.012.

(37) Weon, B. M.; Je, J. H. Fingering inside the Coffee Ring. *Physical Review E* **2013**, *87* (1). https://doi.org/10.1103/PhysRevE.87.013003.

(38) Anyfantakis, M.; Geng, Z.; Morel, M.; Rudiuk, S.; Baigl, D. Modulation of the Coffee-Ring Effect in Particle/Surfactant Mixtures: The Importance of Particle–Interface Interactions. *Langmuir* **2015**, *31* (14), 4113–4120. https://doi.org/10.1021/acs.langmuir.5b00453.

(39) Ryu, S.; Kim, J. Y.; Kim, S. Y.; Weon, B. M. Drying-Mediated Patterns in Colloid-Polymer Suspensions. *Scientific Reports* **2017**, *7* (1). https://doi.org/10.1038/s41598-017-00932-z.

(40) Marín, Á. G.; Gelderblom, H.; Lohse, D.; Snoeijer, J. H. Order-to-Disorder Transition in Ring-Shaped Colloidal Stains. *Physical Review Letters* **2011**, *107* (8). https://doi.org/10.1103/PhysRevLett.107.085502.

(41) Deegan, R. D. Pattern Formation in Drying Drops. *Physical Review E* **2000**, *61* (1), 475–485. https://doi.org/10.1103/PhysRevE.61.475.

(42) Popov, Y. O. Evaporative Deposition Patterns: Spatial Dimensions of the Deposit. *Physical Review E* **2005**, *71* (3). https://doi.org/10.1103/PhysRevE.71.036313.

(43) Pauchard, L.; Parisse, F.; Allain, C. Influence of Salt Content on Crack Patterns Formed through Colloidal Suspension Desiccation. *Physical Review E* **1999**, *59* (3), 3737–3740. https://doi.org/10.1103/PhysRevE.59.3737.

(44) Allain, C.; Limat, L. Regular Patterns of Cracks Formed by Directional Drying of a Collodial Suspension. *Physical Review Letters* **1995**, *74* (15), 2981–2984. https://doi.org/10.1103/PhysRevLett.74.2981.

(45) Giorgiutti-Dauphiné, F.; Pauchard, L. Elapsed Time for Crack Formation during Drying. *The European Physical Journal E* **2014**, *37* (5). https://doi.org/10.1140/epje/i2014-14039-8.

(46) Carle, F.; Brutin, D. How Surface Functional Groups Influence Fracturation in Nanofluid Droplet Dry-Outs. *Langmuir* **2013**, *29* (32), 9962–9966. https://doi.org/10.1021/la401428v.

(47) Anyfantakis, M.; Baigl, D.; Binks, B. P. Evaporation of Drops Containing Silica Nanoparticles of Varying Hydrophobicities: Exploiting Particle–Particle Interactions for Additive-Free Tunable Deposit Morphology. *Langmuir* **2017**, *33* (20), 5025–5036. https://doi.org/10.1021/acs.langmuir.7b00807.

(48) Li, H.; Luo, H.; Zhang, Z.; Li, Y.; Xiong, B.; Qiao, C.; Cao, X.; Wang, T.; He, Y.; Jing, G. Direct Observation of Nanoparticle Multiple-Ring Pattern Formation during Droplet Evaporation with Dark-Field Microscopy. *Physical Chemistry Chemical Physics* **2016**, *18* (18), 13018–13025. https://doi.org/10.1039/C6CP00593D.

(49) Li, Y.; Lv, C.; Li, Z.; Quéré, D.; Zheng, Q. From Coffee Rings to Coffee Eyes. *Soft Matter* **2015**, *11* (23), 4669–4673. https://doi.org/10.1039/C5SM00654F.

(50) Maheshwari, S.; Zhang, L.; Zhu, Y.; Chang, H.-C. Coupling Between Precipitation and Contact-Line Dynamics: Multiring Stains and Stick-Slip Motion. *Physical Review Letters* **2008**, *100* (4). https://doi.org/10.1103/PhysRevLett.100.044503.

(51) Ristenpart, W. D.; Kim, P. G.; Domingues, C.; Wan, J.; Stone, H. A. Influence of Substrate Conductivity on Circulation Reversal in Evaporating Drops. *Physical Review Letters* **2007**, *99* (23). https://doi.org/10.1103/PhysRevLett.99.234502.

**TOC**

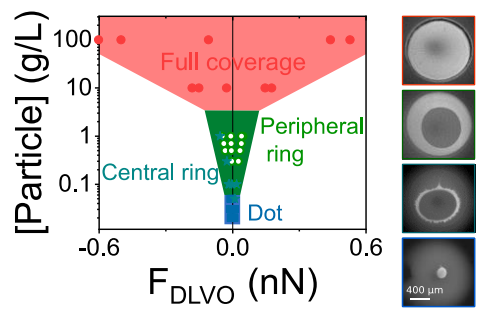